\documentclass[conference]{IEEEtran}

\usepackage[noadjust]{cite}
\usepackage{amsmath,amssymb,amsfonts}
\usepackage{booktabs} % For professional-quality tables
\usepackage{url}      % For formatting URLs
\usepackage{graphicx}
\usepackage{flushend} % Automatically balances columns on the last page

\usepackage{tikz}
\usetikzlibrary{shapes, arrows.meta, positioning, shadows, calc, fit, backgrounds}
\usepackage{pgfplots}
\pgfplotsset{compat=1.18}

\usepackage{hyperref}
\hypersetup{
    colorlinks=true,
    allcolors=black,
    pdfborder={0 0 0},
    pdftitle={SHERLOCK: A Deep Learning Approach To Detect Software Vulnerabilities},
    pdfauthor={Saadh Jawwadh}
}

\newcommand{\MYhyperreffix}{}
\let\MYhyperreffix\cite
\renewcommand{\cite}[1]{\MYhyperreffix{#1}}

\tikzset{
    layerbox/.style={draw, rounded corners, inner sep=10pt, align=center, drop shadow, fill=white, dashed},
    process/.style={rectangle, minimum width=2.2cm, minimum height=1cm, text centered, draw=black, fill=white, rounded corners, font=\small},
    storage/.style={cylinder, cylinder uses custom fill, cylinder body fill=white, cylinder end fill=gray!20, shape border rotate=90, aspect=0.25, draw, minimum width=1.2cm, minimum height=1.2cm, align=center, font=\small},
    arrow/.style={thick, ->, >=Stealth},
    cnn_layer/.style={draw, rectangle, minimum width=3.5cm, minimum height=0.7cm, align=center, fill=blue!5, rounded corners, font=\footnotesize},
    cnn_output/.style={draw, rectangle, minimum width=1cm, minimum height=0.6cm, align=center, fill=red!5, font=\scriptsize}
}

\title{SHERLOCK: A Deep Learning Approach To Detect Software Vulnerabilities}

\author{
    \IEEEauthorblockN{Saadh Jawwadh}
    \IEEEauthorblockA{\textit{Informatics Institute of Technology} \\
    in collaboration with \textit{University of Westminster}\\
    Colombo, Sri Lanka \\}
    \and
    \IEEEauthorblockN{Guhanathan Poravi}
    \IEEEauthorblockA{\textit{Supervisor} \\
    \textit{Informatics Institute of Technology} \\
    Colombo, Sri Lanka}
}

\begin{document}

\maketitle
\thispagestyle{plain}
\pagestyle{plain}

% --- ABSTRACT ---
\begin{abstract}
The increasing reliance on software in various applications has made the problem of software vulnerability detection more critical. Software vulnerabilities can lead to security breaches, data theft, and other negative outcomes. Traditional software vulnerability detection techniques, such as static and dynamic analysis, have been shown to be ineffective at detecting multiple vulnerabilities.

To address this issue, this study employed a deep learning approach, specifically Convolutional Neural Networks (CNN), to solve the software vulnerability detection problem. A 5-split cross-validation approach was used to train and evaluate the CNN model, which takes tokenized source code as input.

The findings indicated that Sherlock successfully detected multiple vulnerabilities at the function level, and its performance was particularly strong for CWE-199, CWE-120, and CWE-Other, with an overall high accuracy rate and significant true positive and true negative values. However, the performance was less reliable for some vulnerabilities due to the lack of a standardized dataset which will be a future research direction. The results suggest that compared to current techniques, the proposed deep learning approach has the potential to substantially enhance the accuracy of software vulnerability detection.
\end{abstract}

\begin{IEEEkeywords}
Software Vulnerability Detection, AI, Deep Learning, Convolutional Neural Network, Gaussian Noise
\end{IEEEkeywords}

% --- I. INTRODUCTION ---
\section{Introduction}
Software vulnerability is a critical security flaw or weakness in software code that an attacker can exploit \cite{hanif_rise_2021}. The rapid expansion of interconnected computer systems has led to a proportional increase in these vulnerabilities, causing significant financial losses and downtime \cite{hanif_rise_2021, skybox_security_vulnerability_2022, zero_day_initiative_zero_2023}. Software vulnerabilities can have a drastic impact on organizations and individuals, including financial losses, denial of service, reputational damage, data loss, and legal issues \cite{tomaschek_lastpass_2023, coker_lastpass_2023, aiyer_new_2022}. This dramatic rise necessitates more effective and efficient detection methods \cite{noauthor_state_2021}.

Current detection techniques are broadly categorized as static, dynamic, and hybrid. Static analysis (e.g., rule-based analysis, symbolic execution) analyzes source code without execution but suffers from high false positive rates \cite{lin_software_2020, cowan_stackguard_1998, pewny_leveraging_2014}. Dynamic analysis (e.g., fuzzy testing, taint analysis) examines software during runtime but often has low code coverage \cite{newsome_dynamic_2005}. Hybrid approaches attempt to combine the two but have their own limitations \cite{yamaguchi_modeling_2014}.

The ineffectiveness of these traditional methods has led to new investigations into data-driven, machine learning-based approaches, which have shown promising results \cite{lin_software_2020, sun_data-driven_2019, coulter_data-driven_2020}. However, despite advancements in AI, its application to software vulnerability detection is relatively under-researched \cite{ghaffarian_software_2018, sun_data-driven_2019}. Software vulnerabilities have become a widespread issue for the modern generation, and exploitable vulnerabilities can pose a threat to computer systems \cite{noauthor_heartbleed_2020}. A significant research gap exists for solutions that can detect \textit{multiple} vulnerabilities \textit{prior} to deployment, as most current solutions focus on single, known vulnerabilities post-deployment \cite{sonnekalb_deep_2021, ryan_project_2022, singh_cyber_2022}.

This paper introduces SHERLOCK, a novel deep learning system designed to address this gap. The aim of this research is to design, develop, and evaluate an AI-based system capable of identifying multiple software vulnerabilities at the function-level from source code. We hypothesize that a Convolutional Neural Network (CNN) model, trained on a large dataset of tokenized code, can significantly improve detection accuracy over existing methods.

% --- II. RELATED WORK ---
\section{Related Work}
The state-of-the-art in vulnerability detection has shifted from purely manual or static analysis to more sophisticated data-driven techniques.

\subsection{Code Analysis-Based Detection}
Traditional methods like static and dynamic analysis form the baseline for vulnerability detection. While fundamental, they are often inefficient when faced with the massive recent spike in vulnerability reports. They are time-consuming, require significant expertise, and suffer from either high false positives (static) or low code coverage (dynamic) \cite{ghaffarian_software_2018, lin_software_2020, hanif_rise_2021}.

\subsection{Data-Driven Detection}
Data-driven approaches, using machine learning and deep learning, have gained significant attention \cite{russell_automated_2018}. These methods use large datasets to train models to identify patterns and anomalies indicative of vulnerabilities \cite{ghaffarian_software_2018}. They have the potential for higher accuracy, better code coverage, and reduced false positives. However, this field faces its own challenges, primarily the scarcity of large, high-quality, labeled datasets \cite{lin_software_2020, ghaffarian_software_2018, bilgin_vulnerability_2020}.

\subsection{Significant Deep Learning Works}
Several key studies form the foundation for this work. \textbf{VulDeePecker} \cite{li_vuldeepecker_2018} was a pioneering deep learning-based system that used a Bidirectional LSTM (BLSTM) on tokenized code. However, it focused on slice-level extraction for only two vulnerability types.

A foundational work by \textbf{Russell et al.} \cite{russell_automated_2018} was the first to prove that deep learning, particularly CNNs, could outperform traditional machine learning (e.g., Random Forest) for function-level vulnerability detection. They also released a large, labeled dataset of 1.2 million C/C++ functions (the Draper VDISC dataset), which has become a standard for this research domain. Their work, however, focused on a binary (vulnerable/not-vulnerable) classification rather than identifying \textit{types} of vulnerabilities.

Other works, such as SySeVR \cite{li_sysevr_2022} and a study by Bilgin et al. \cite{bilgin_vulnerability_2020}, have explored different representations (like Abstract Syntax Trees) and models (like MLPs), but CNNs remain a consistently strong performer for this task. SHERLOCK builds upon the function-level CNN approach of Russell et al. \cite{russell_automated_2018} but extends it to solve the multi-vulnerability classification problem.

% --- III. THE SHERLOCK METHODOLOGY ---
\section{The SHERLOCK Methodology}
\begin{figure}[htbp]
\centering
% If the filename contains spaces, surround it with double quotes as below.
% Alternatively, rename the file to remove spaces (recommended) and update the filename.
\includegraphics[width=0.95\columnwidth,keepaspectratio]{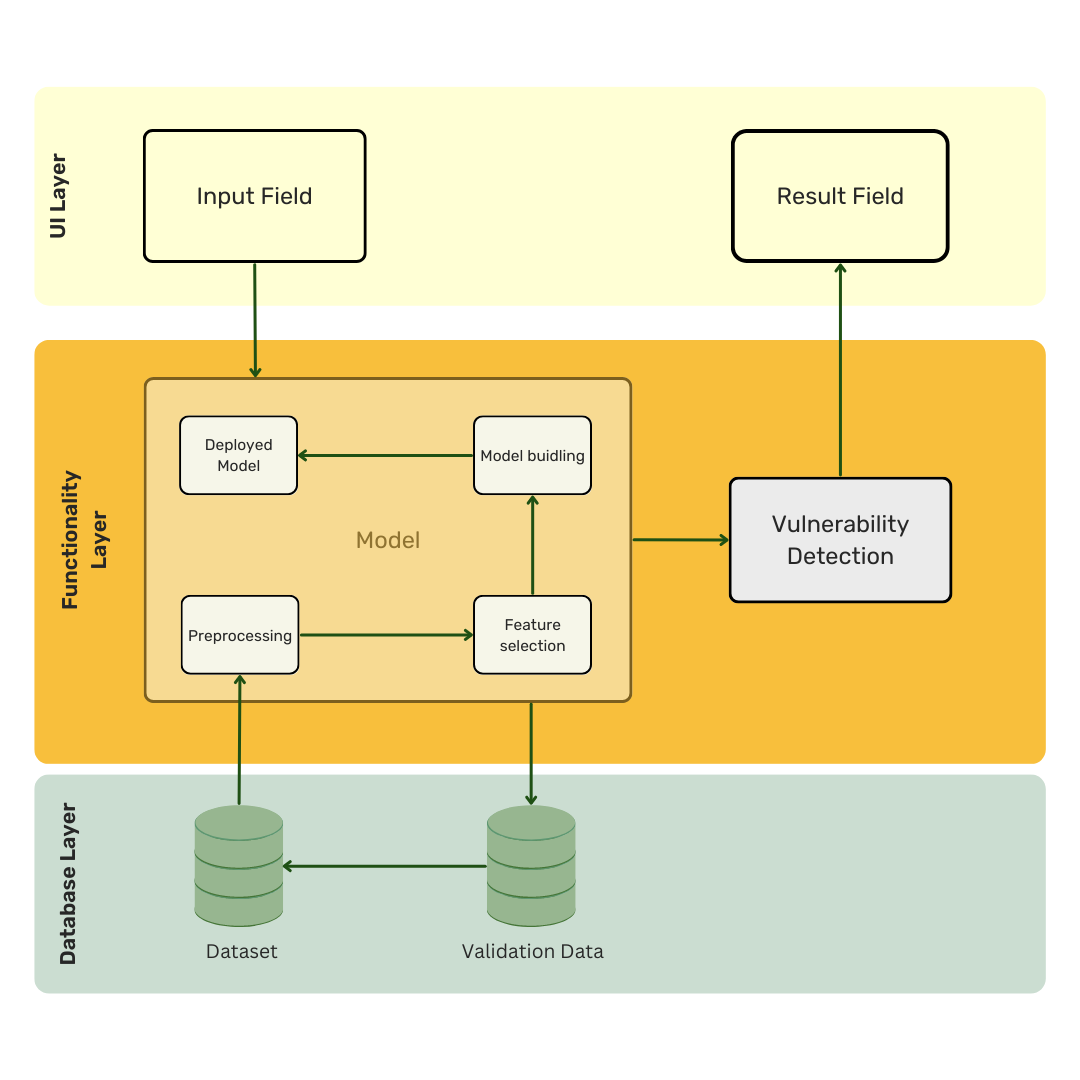}
\caption{The Layered Architecture of SHERLOCK. The system separates the User Interface, Core Functionality (AI Model), and Data Management into distinct layers for modularity and scalability.}
\label{fig:architecture}
\end{figure}

SHERLOCK is designed as a three-layer system (see Fig. \ref{fig:architecture}) that processes raw source code, analyzes it using a deep learning model, and presents a multi-class vulnerability report to the user.

\subsection{System Architecture}
The system is comprised of three distinct layers:
\begin{itemize}
    \item \textbf{UI Layer:} A user interface (implemented as a web application) that allows a user to input C/C++ source code.
    \item \textbf{Functionality Layer:} The core of the system. It contains the data preprocessing pipeline and the deployed CNN model. It receives code from the UI layer, tokenizes it, and feeds it to the model for inference.
    \item \textbf{Database Layer:} This layer contains the dataset (Draper VDISC) used for training the model and the validation data used for evaluation.
\end{itemize}

% --- FIGURE 1: ARCHITECTURE (PNG) ---

\subsection{Dataset}
We use the \textbf{Draper VDISC dataset} \cite{russell_automated_2018}, which contains 1.27 million C/C++ functions from open-source software. The dataset is labeled for five vulnerability categories: CWE-120 (Buffer Copy without Checking Size), CWE-119 (Improper Restriction of Operations within bounds of memory buffer), CWE-469 (NULL Pointer Dereference), CWE-476 (NULL Pointer Dereference), and a composite "CWE-Other" category. The data is stored in HDF5 files, with each function's source code as a string. We use an 80\%/10\%/10\% split for training, validation, and testing, respectively.

\subsection{Model Implementation}
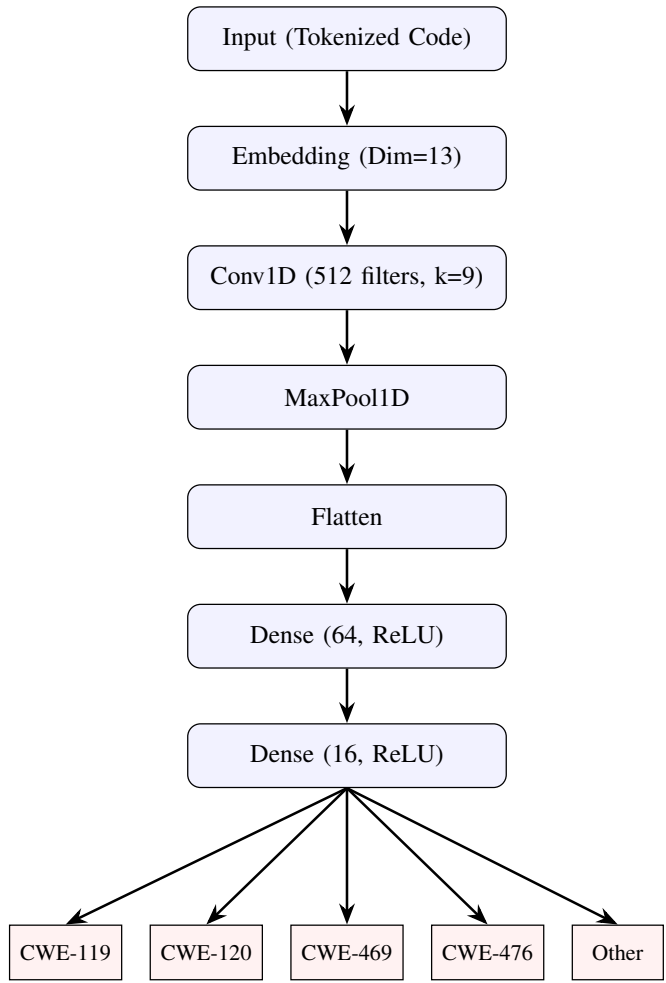
\begin{figure}[htbp]
\centering
\resizebox{\columnwidth}{!}{%
\begin{tikzpicture}[node distance=0.6cm]
    % Main Layers Stack
    \node[cnn_layer] (input) {Input (Tokenized Code)};
    \node[cnn_layer, below=of input] (embed) {Embedding (Dim=13)};
    \node[cnn_layer, below=of embed] (conv) {Conv1D (512 filters, k=9)};
    \node[cnn_layer, below=of conv] (pool) {MaxPool1D};
    \node[cnn_layer, below=of pool] (flat) {Flatten};
    \node[cnn_layer, below=of flat] (dense1) {Dense (64, ReLU)};
    \node[cnn_layer, below=of dense1] (dense2) {Dense (16, ReLU)};

    % Output Heads (Fan Out)
    \node[cnn_output, below=1.5cm of dense2] (out3) {CWE-469};
    \node[cnn_output, left=0.3cm of out3] (out2) {CWE-120};
    \node[cnn_output, left=0.3cm of out2] (out1) {CWE-119};
    \node[cnn_output, right=0.3cm of out3] (out4) {CWE-476};
    \node[cnn_output, right=0.3cm of out4] (out5) {Other};

    % Connections
    \draw[arrow] (input) -- (embed);
    \draw[arrow] (embed) -- (conv);
    \draw[arrow] (conv) -- (pool);
    \draw[arrow] (pool) -- (flat);
    \draw[arrow] (flat) -- (dense1);
    \draw[arrow] (dense1) -- (dense2);

    % Fan out connections
    \foreach \x in {out1, out2, out3, out4, out5}
        \draw[arrow] (dense2.south) -- (\x.north);

\end{tikzpicture}
}
\caption{CNN Model Architecture. The model processes tokenized code through convolutional layers before splitting into five separate output heads, allowing for multi-label classification.}
\label{fig:model}
\end{figure}

The core of SHERLOCK is a Convolutional Neural Network (CNN) built using TensorFlow and Keras. The data pipeline and model architecture are as follows (see Fig. \ref{fig:model}):

\begin{enumerate}
    \item \textbf{Preprocessing \& Tokenization:} The raw C/C++ function source code is tokenized. This process converts the string of code into a sequence of integer tokens, which can be fed into a neural network.

    \item \textbf{Embedding Layer:} The sequence of tokens is passed to an \texttt{Embedding} layer. This layer maps each integer token to a dense vector of a fixed size (13 dimensions in our case). This allows the model to learn a semantic representation for each token in the code.

    \item \textbf{Convolutional Layer:} A 1D Convolutional layer (\texttt{Convolution1D}) with 512 filters and a kernel size of 9 is applied. This layer acts as a feature extractor, learning to identify significant patterns (n-grams) in the sequence of code tokens that may indicate a vulnerability. A \texttt{ReLU} activation function is used.

    \item \textbf{Pooling \& Regularization:} A \texttt{MaxPool1D} layer is used to down-sample the feature maps, retaining the most important features. A \texttt{Dropout} layer (0.5) is applied to prevent overfitting.

    \item \textbf{Dense Layers:} The features are flattened and passed through two \texttt{Dense} (fully-connected) layers (64 and 16 neurons, respectively, with \texttt{ReLU} activation) to perform high-level feature combination.

    \item \textbf{Multi-Output Head:} The key innovation of SHERLOCK is its multi-output head. Instead of a single binary classification, the final dense layer feeds into five separate 2-neuron \texttt{Dense} output layers, one for each vulnerability class (CWE-199, CWE-120, etc.). Each output layer uses a \texttt{softmax} activation function to produce a probability distribution (vulnerable/not-vulnerable) for its specific CWE.
\end{enumerate}

The model is compiled with a custom \texttt{Adam} optimizer (learning rate 0.005) and uses \texttt{categorical\_crossentropy} as the loss function, as each head is a categorical classifier.

% --- FIGURE 2: CNN MODEL (TikZ) ---

% --- IV. EVALUATION AND RESULTS ---
\section{Evaluation and Results}
We evaluated SHERLOCK's performance on the 10\% test split of the VDISC dataset. The primary goal was to assess its ability to accurately classify functions for each of the five vulnerability types.

\subsection{Performance Metrics}
We used standard classification metrics: Accuracy, Precision, Recall, F1-Score, and Area Under the Curve (AUC).

\begin{itemize}
    \item \textbf{True Negatives (TN)} were high across all categories. The model is excellent at correctly identifying non-vulnerable code.
    \item \textbf{True Positives (TP)} were strong for CWE-199, CWE-120, and CWE-Other. However, the model performed poorly on CWE-469 and CWE-476, identifying almost no true positives.
    \item \textbf{The reason for this poor performance} on specific classes is the extreme class imbalance in the training data. The dataset contains very few positive examples for CWE-469 and CWE-476, making it difficult for the model to learn their patterns.
\end{itemize}

% --- FIGURE 3: PERFORMANCE CHART (PGFPlots) ---
\begin{figure}[htbp]
\centering
\begin{tikzpicture}
\begin{axis}[
    ybar,
    bar width=8pt,
    width=\columnwidth,
    height=6cm,
    ymin=0, ymax=1.15, % Increased ymax for labels
    ylabel={Score},
    symbolic x coords={CWE-199, CWE-120, CWE-469, CWE-476, Other},
    xtick=data,
    xticklabel style={font=\scriptsize, rotate=0},
    legend style={at={(0.5,-0.25)}, anchor=north, legend columns=-1, font=\scriptsize},
    grid=major,
    nodes near coords,
    nodes near coords style={font=\tiny, rotate=90, anchor=west},
    enlarge x limits=0.15,
]
    % Precision
    \addplot coordinates {(CWE-199,0.22) (CWE-120,0.15) (CWE-469,0.00) (CWE-476,0.03) (Other,0.04)};
    % Recall
    \addplot coordinates {(CWE-199,0.17) (CWE-120,0.21) (CWE-469,0.00) (CWE-476,0.03) (Other,0.04)};
    % F1
    \addplot coordinates {(CWE-199,0.19) (CWE-120,0.18) (CWE-469,0.00) (CWE-476,0.03) (Other,0.04)};

    \legend{Precision, Recall, F1-Score}
\end{axis}
\end{tikzpicture}
\caption{Model Performance by Vulnerability Class. While detection is promising for CWE-199 and CWE-120, the low scores for CWE-469 and CWE-476 highlight the challenge of data imbalance.}
\label{fig:performance_chart}
\end{figure}
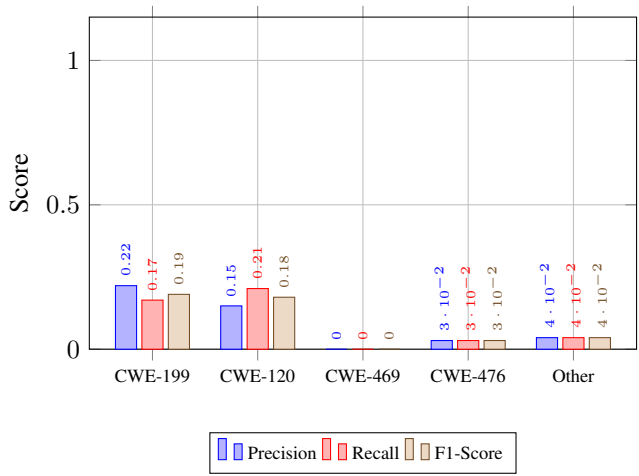

\subsection{Quantitative Results}
Table \ref{tab:performance} shows the detailed metrics for each vulnerability. The model achieves high accuracy (92\%-99\%) and good AUC scores (0.72-0.83) for the three classes it successfully learned (CWE-199, CWE-120, CWE-Other), indicating a strong and reliable classification ability. The metrics for CWE-469 and CWE-476 reflect the failure to learn from the imbalanced data.

\begin{table}[ht]
\centering
\caption{Model Performance Metrics}
\label{tab:performance}
\resizebox{\columnwidth}{!}{%
\begin{tabular}{@{}lccccc@{}}
\toprule
\textbf{Metric (CWE)} & \textbf{Accuracy} & \textbf{Precision} & \textbf{Recall} & \textbf{F1 Score} & \textbf{AUC} \\
\midrule
CWE-199               & 0.97              & 0.22               & 0.17            & 0.19              & 0.81         \\
CWE-120               & 0.92              & 0.15               & 0.21            & 0.18              & 0.72         \\
CWE-469               & 0.99              & 0.00               & 0.00            & 0.00              & 0.83         \\
CWE-476               & 0.98              & 0.03               & 0.03            & 0.03              & 0.54         \\
CWE-Other             & 0.95              & 0.04               & 0.04            & 0.04              & 0.67         \\
\bottomrule
\end{tabular}
}
\end{table}

\subsection{Benchmarking}
We benchmarked SHERLOCK against a baseline model (Code2vec + MLP) from a similar study \cite{bilgin_vulnerability_2020}, focusing on the CWE-199 class for a direct comparison. As shown in Table \ref{tab:benchmarking}, SHERLOCK achieves a significantly higher F1-Score (0.19 vs 0.12) and Precision (0.22 vs 0.06), though at the cost of lower Recall. This demonstrates that our CNN-based approach is competitive and, in terms of precision, superior to the baseline.

\begin{table}[ht]
\centering
\caption{Benchmarking vs. Baseline Model (CWE-199)}
\label{tab:benchmarking}
\begin{tabular}{@{}lccc@{}}
\toprule
\textbf{Model}                & \textbf{Precision} & \textbf{Recall} & \textbf{F1 Score} \\
\midrule
Code2vec + MLP \cite{bilgin_vulnerability_2020} & 0.06               & 0.87            & 0.12              \\
\textbf{Sherlock (Ours)}      & \textbf{0.22}      & \textbf{0.17}   & \textbf{0.19}     \\
\bottomrule
\end{tabular}
\end{table}

% --- V. CONCLUSION AND FUTURE WORK ---
\section{Conclusion and Future Work}
This research successfully designed, developed, and evaluated SHERLOCK, a novel deep learning system for detecting multiple software vulnerabilities from source code. The aim of the research was achieved, demonstrating that a CNN-based approach with a multi-output head can effectively identify multiple vulnerability types (CWE-199, CWE-120, CWE-Other) with high accuracy and reliability.

The primary limitation of this research is the severe class imbalance in the Draper VDISC dataset. This imbalance prevented the model from effectively learning to detect CWE-469 and CWE-476.

Future enhancements for SHERLOCK will focus on three main areas:
\begin{enumerate}
    \item \textbf{Addressing Data Imbalance:} Create a more balanced, labeled dataset, potentially using data augmentation or synthetic data generation techniques (e.g., resampling) to improve performance on rare vulnerability classes.
    \item \textbf{Expanding Language Support:} Train the model on source code from other programming languages, such as Python or Java, to increase its applicability \cite{zhou_large_2024}.
    \item \textbf{Alternative Architectures:} Experiment with ensemble models or NLP-based transformers (e.g., BERT) to potentially improve feature extraction and overall detection performance \cite{salem_advancing_2024}.
\end{enumerate}

In conclusion, SHERLOCK is a promising proof-of-concept that demonstrates the potential of deep learning to move beyond single-binary classification and toward a more practical, multi-vulnerability detection paradigm, ultimately helping developers write more secure code prior to deployment.

% ----------------------------------------------------------------------
% --- REFERENCES ---
% ----------------------------------------------------------------------
\bibliographystyle{IEEEtran}
\bibliography{SVD}

\end{document}